\documentclass[twocolumn,floatfix]{aastex631}
\usepackage{amsmath}
\usepackage{pgfplots}
\pgfplotsset{compat=1.16}
\usepackage{tikzscale}
\usepackage{censor}
\usepackage{xfrac}
\usepackage{booktabs}
\usepackage{microtype}
\usepackage[caption=false, font=small]{subfig}
\usepackage{makecell}
\let\tablenum\relax
\usepackage{siunitx}
\usepackage[noabbrev,capitalize]{cleveref}
\usepackage{glossaries-extra}
\glsdisablehyper
\usepackage{etoolbox}

\usepackage{silence}
\makeatletter
\patchcmd\H@refstepcounter{\protected@edef}{\protected@xdef}{}{}
\makeatother

\makeatletter
\patchcmd\linenumberpar{\@LN@parpgbrk}{\penalty\@LN@parpgpen\relax}{}{}
\makeatother

\definecolor{darklavender}{rgb}{0.45, 0.31, 0.59}
\definecolor{light-gray}{rgb}{0.8,0.8,0.8}
\def\censorcolor{light-gray}
\let\svcensorrule\censorrule
\renewcommand\censorrule[1]{%
    \textcolor{\censorcolor}{\svcensorrule{#1}}}

\newcommand{\gsfcAffiliationString}{NASA Goddard Space Flight Center, Greenbelt, MD 20771, USA}
\newcommand{\cuaAffiliationString}{Department of Physics, The Catholic University of America, Washington, DC 20064, USA}
\newcommand{\umdAffiliationString}{Department of Astronomy, University of Maryland, College Park, MD 20742, USA}

\newcommand{\zahAffiliationString}{Zentrum f{\"u}r Astronomie der Universit{\"a}t Heidelberg, Astronomisches Rechen-Institut, M{\"o}nchhofstr.\ 12-14, 69120 Heidelberg, Germany}
\newcommand{\orauAffiliationString}{Oak Ridge Associated Universities, Oak Ridge, TN 37830, USA}

\newcommand{\moaAffiliationString}{The MOA collaboration}

\newcommand*\diff{\mathop{}\!\mathrm{d}}

\let\emph\relax 
\DeclareTextFontCommand{\emph}{\bfseries\em}

\begin{document}

\title{MOA-2020-BLG-135Lb: A New Neptune-class Planet for the Extended MOA-II Exoplanet Microlens Statistical Analysis}

\author[0000-0003-2267-1246]{Stela {Ishitani~Silva}}
\affiliation{\gsfcAffiliationString}
\affiliation{\cuaAffiliationString}
\affiliation{\moaAffiliationString}
\author[0000-0003-2388-4534]{Cl{\'e}ment Ranc}
\affiliation{\zahAffiliationString}
\affiliation{\moaAffiliationString}
\author[0000-0001-8043-8413]{David P. Bennett}
\affiliation{\gsfcAffiliationString}
\affiliation{\umdAffiliationString}
\affiliation{\moaAffiliationString}
\author{Ian A.~Bond}
\affiliation{Institute of Natural and Mathematical Sciences, Massey University, Auckland 0745, New Zealand}
\affiliation{\moaAffiliationString}
\author[0000-0001-6000-3463]{Weicheng Zang}
\affiliation{Department of Astronomy, Tsinghua University, Beijing 100084, China}
\affiliation{The CFHT Microlensing Collaboration}
\collaboration{5}{(Leading authors)}

\author{Fumio Abe}
\affiliation{Institute for Space-Earth Environmental Research, Nagoya University, Nagoya 464-8601, Japan}
\author{Richard K. Barry}
\affiliation{\gsfcAffiliationString}
\author{Aparna Bhattacharya}
\affiliation{\gsfcAffiliationString}
\affiliation{\umdAffiliationString}
\author{Hirosane Fujii}
\affiliation{Institute for Space-Earth Environmental Research, Nagoya University, Nagoya 464-8601, Japan}
\author{Akihiko Fukui}
\affiliation{Department of Earth and Planetary Science, Graduate School of Science, The University of Tokyo, 7-3-1 Hongo, Bunkyo-ku, Tokyo 113-0033, Japan}
\affiliation{Instituto de Astrof\'isica de Canarias, V\'ia L\'actea s/n, E-38205 La Laguna, Tenerife, Spain}
\author{Yuki Hirao}
\affiliation{Department of Earth and Space Science, Graduate School of Science, Osaka University, Toyonaka, Osaka 560-0043, Japan}
\author{Yoshitaka Itow}
\affiliation{Institute for Space-Earth Environmental Research, Nagoya University, Nagoya 464-8601, Japan}
\author{Rintaro Kirikawa}
\affiliation{Department of Earth and Space Science, Graduate School of Science, Osaka University, Toyonaka, Osaka 560-0043, Japan}
\author{Iona Kondo}
\affiliation{Department of Earth and Space Science, Graduate School of Science, Osaka University, Toyonaka, Osaka 560-0043, Japan}
\author[0000-0003-2302-9562]{Naoki Koshimoto}
\affiliation{\gsfcAffiliationString}
\affiliation{\umdAffiliationString}
\affiliation{Department of Earth and Space Science, Graduate School of Science, Osaka University, Toyonaka, Osaka 560-0043, Japan}
\author{Yutaka Matsubara}
\affiliation{Institute for Space-Earth Environmental Research, Nagoya University, Nagoya 464-8601, Japan}
\author{Sho Matsumoto}
\affiliation{Department of Earth and Space Science, Graduate School of Science, Osaka University, Toyonaka, Osaka 560-0043, Japan}
\author{Shota Miyazaki}
\affiliation{Department of Earth and Space Science, Graduate School of Science, Osaka University, Toyonaka, Osaka 560-0043, Japan}
\author{Yasushi Muraki}
\affiliation{Institute for Space-Earth Environmental Research, Nagoya University, Nagoya 464-8601, Japan}
\author[0000-0001-8472-2219]{Greg Olmschenk}
\affiliation{\gsfcAffiliationString}
\affiliation{\orauAffiliationString}
\author{Arisa Okamura}
\affiliation{Department of Earth and Space Science, Graduate School of Science, Osaka University, Toyonaka, Osaka 560-0043, Japan}
\author{Nicholas J. Rattenbury}
\affiliation{Department of Physics, University of Auckland, Private Bag 92019, Auckland, New Zealand}
\author{Yuki Satoh}
\affiliation{Department of Earth and Space Science, Graduate School of Science, Osaka University, Toyonaka, Osaka 560-0043, Japan}
\author{Takahiro Sumi}
\affiliation{Department of Earth and Space Science, Graduate School of Science, Osaka University, Toyonaka, Osaka 560-0043, Japan}
\author{Daisuke Suzuki}
\affiliation{Department of Earth and Space Science, Graduate School of Science, Osaka University, Toyonaka, Osaka 560-0043, Japan}
\author{Taiga Toda}
\affiliation{Department of Earth and Space Science, Graduate School of Science, Osaka University, Toyonaka, Osaka 560-0043, Japan}
\author{Paul . J. Tristram}
\affiliation{University of Canterbury Mt.\ John Observatory, P.O. Box 56, Lake Tekapo 8770, New Zealand}
\author{Aikaterini Vandorou}
\affiliation{\gsfcAffiliationString}
\affiliation{\umdAffiliationString}
\author{Hibiki Yama}
\affiliation{Department of Earth and Space Science, Graduate School of Science, Osaka University, Toyonaka, Osaka 560-0043, Japan}
\collaboration{24}{(The MOA Collaboration)}

\author{Andreea Petric}
\affiliation{CFHT Corporation, 65-1238 Mamalahoa Hwy, Kamuela, Hawaii 96743, USA}
\affiliation{Space Telescope Science Institute, Baltimore, MD 21211, USA}
\author{Todd Burdullis}
\affiliation{CFHT Corporation, 65-1238 Mamalahoa Hwy, Kamuela, Hawaii 96743, USA}
\author{Pascal Fouqu\'e}
\affiliation{CFHT Corporation, 65-1238 Mamalahoa Hwy, Kamuela, Hawaii 96743, USA}
\affiliation{Universit\'e de Toulouse, UPS-OMP, IRAP, Toulouse, France}
\author{Shude Mao}
\affiliation{Department of Astronomy, Tsinghua University, Beijing 100084, China}
\affiliation{National Astronomical Observatories, Chinese Academy of Sciences, Beijing 100101, China}
\author[0000-0001-7506-5640]{Matthew T. Penny}
\affiliation{Department of Astronomy, The Ohio State University, 140 W. 18th Avenue, Columbus, OH 43210, USA}
\author{Wei Zhu}
\affiliation{Department of Astronomy, Tsinghua University, Beijing 100084, China}
\collaboration{6}{(The CFHT Microlensing Collaboration)}  
\author[0000-0002-3042-4539]{Gioia Rau}
\affiliation{\gsfcAffiliationString}
\affiliation{\cuaAffiliationString}
\nocollaboration{1}

\begin{abstract}
We report the light-curve analysis for the event MOA-2020-BLG-135, which leads to the discovery of a new Neptune-class planet, MOA-2020-BLG-135Lb. With a derived mass ratio of $q=1.52_{-0.31}^{+0.39} \times 10^{-4}$ and separation $s\approx1$, the planet lies exactly at the break and likely peak of the exoplanet mass-ratio function derived by the MOA collaboration \citep{suzuki2016exoplanet}. We estimate the properties of the lens system based on a Galactic model and considering two different Bayesian priors: one assuming that all stars have an equal planet-hosting probability and the other that planets are more likely to orbit more massive stars. With a uniform host mass prior, we predict that the lens system is likely to be a planet of mass  $m_\mathrm{planet}=11.3_{-6.9}^{+19.2} M_\oplus$ and a host star of mass $M_\mathrm{host}=0.23_{-0.14}^{+0.39} M_\odot$, located at a distance $D_L=7.9_{-1.0}^{+1.0}\;\mathrm{kpc}$. With a prior that holds that planet  occurrence scales in proportion to the host star mass, the estimated lens system properties are $m_\mathrm{planet}=25_{-15}^{+22} M_\oplus$, $M_\mathrm{host}=0.53_{-0.32}^{+0.42} M_\odot$, and $D_L=8.3_{-1.0}^{+0.9}\; \mathrm{kpc}$. This planet qualifies for inclusion in the extended MOA-II exoplanet microlens sample.
\end{abstract}

\section{Introduction}

Gravitational microlensing \citep{mao1991gravitational} has been solidified as one of the main techniques for detecting planets, being most sensitive to low-mass planets \citep{bennett1996detecting} that orbit at moderate to large distances from their host star \citep{gould1992discovering}, typically from $0.5\text{-}10\,\text{AU}$, complementing other exoplanet detection methods \citep{bennett2008exoplanets,gaudi2012microlensing, batista2018finding,guerrero2021tess}. The first planetary microlensing event was discovered in 2004 \citep{bond2004ogle}, and since then, more than 120 exoplanets have been discovered by the method of gravitational microlensing. 

The state-of-the-art statistical analysis of planetary signals discovered using gravitational microlensing, \citet{suzuki2016exoplanet}, implied that cold Neptunes were likely to be the most common type of planets beyond the snow line. This inference was done by discovering a break and likely peak in the planet–to–host star mass-ratio function for a mass ratio $q\sim10^{-4}$ when studying the MOA-II microlensing events from 2007 to 2012. At the time of that statistical analysis, it was possible to conclude that while the \citet{suzuki2016exoplanet} sample generally supported the predictions for the planet distribution from core accretion theory population synthesis models for planets beyond the snow line \citep{ida2004toward,mordasini2009extrasolar}, the existence of this Neptune peak in the sample distribution actually added contradictions. These previous models for the planet distribution predicted the existence of a sub-Saturn mass planet desert, which conflicted with the microlensing observations \citep{suzuki2016exoplanet,suzuki2018microlensing}. Only this year \citet{ali2022effect} published their investigation into the origins of cold sub-Saturns, which concluded that these exoplanets may be more common than what was previously predicted. The recent study by \citet{zang2022systematic} obtained a uniform distribution in $\log q$, which also indicates the absence of a planetary desert of sub-Saturn mass. It is important to notice that the \citet{suzuki2016exoplanet} sample had 30 exoplanets, while \citet{zang2022systematic} had only 13 planets. The size of the samples makes their expansion crucial for future investigations and, consequently, for a better understanding of the distribution of exoplanets.

In this paper, we present the analysis of the gravitational microlensing event MOA-2020-BLG-135 with a short-term planetary lensing signal in its light curve. This analysis leads us to the discovery of MOA-2020-BLG-135Lb, a new planet detected by MOA-II. This new exoplanet qualifies to be included in the upcoming statistical analysis of cold exoplanets detected by the MOA-II survey, which is the expansion of the \citet{suzuki2016exoplanet} sample analysis. This paper, presenting a complete study of this event, is organized as follows. First, we describe the observation of the event in Section~\ref{sec:data_photometry}, and how we obtain our best-fit models and explore the full parameter space in Section \ref{sec:lightcurve_models}. Then, we explain the photometric calibration and how we retrieve the source size in Section~\ref{sec:photometric_calibration_source_radius}, and present our methodology to estimate the lens's physical properties in Section~\ref{sec:lens_system_properties}, and discuss them comparing with previous literature in Section~\ref{sec:discussion}. Finally, we conclude summarizing our results in Section~\ref{sec:conclusion}.

\section{Observations and Data}\label{sec:data_photometry}

\begin{figure*}
    \includegraphics[width=\linewidth]{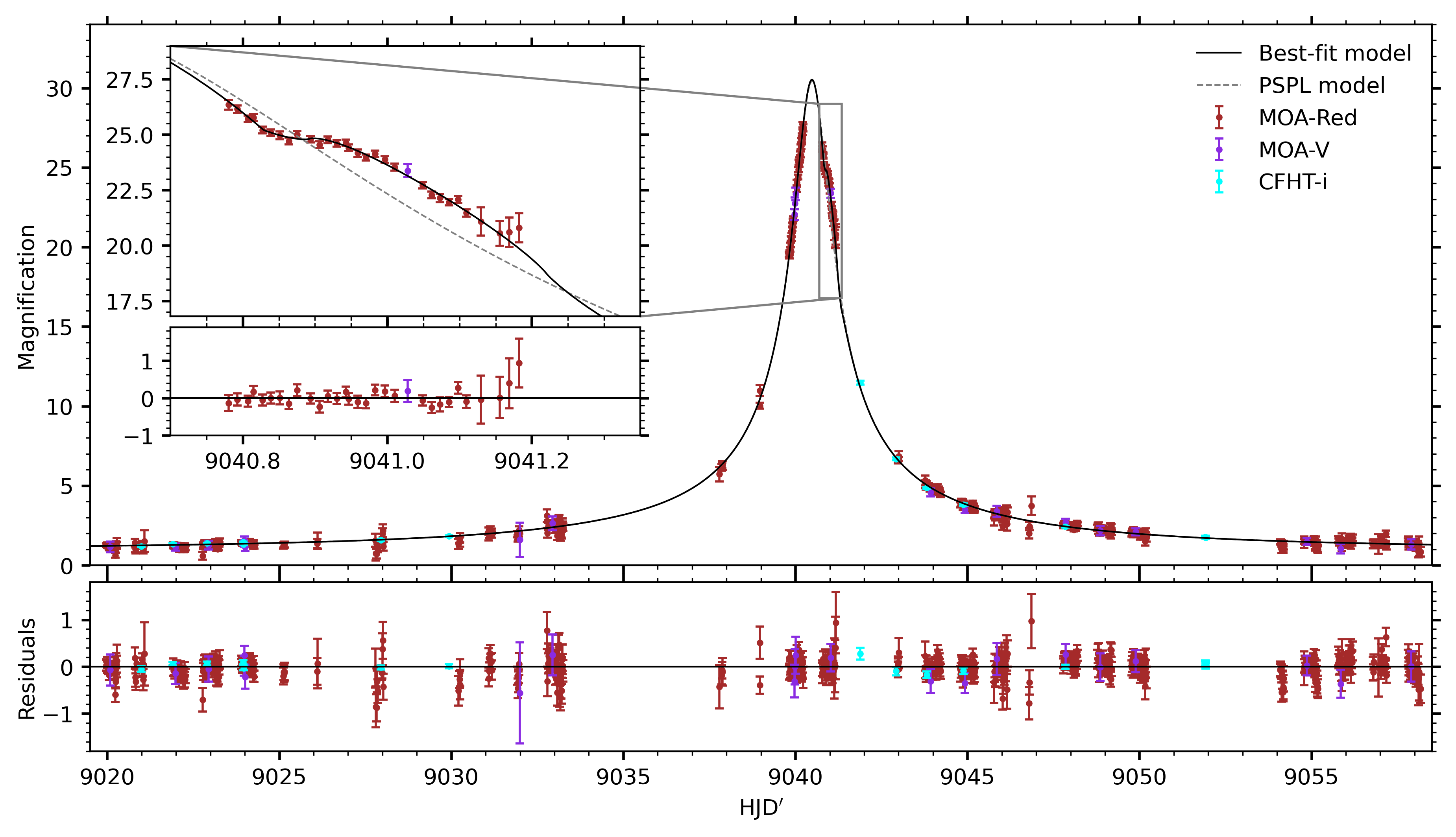} 
    \caption{The best-fit model for the MOA-2020-BLG-135 light curve in a magnification vs. time ($\mathrm{HJD'} = \mathrm{HJD} - 2,450,000$) plot. The MOA-Red, MOA-\textit{V}, and CFHT-\textit{i} data are shown in brown, violet, and blue, respectively. The best single source planetary model 2L1S (i.e., two lenses and one source) $s>1$ is our best-fit model (see Section~\ref{sec:lightcurve_models}) and is displayed as a black solid line, while the Point-Source Point-Lens (PSPL) model is displayed as a dashed gray line. The main panel shows all the photometric data sets overlaid with the best-fit model. The upper left side of the Figure, zooming near the peak of the event, shows the planetary perturbation, which is well covered by the MOA-Red data set, and the corresponding residuals. The lower panel shows the residuals of the best-fit model for each instrument - see colored legend.} 
    \label{fig:bestfit_model}
\end{figure*}

The microlensing event MOA-2020-BLG-135 was discovered by the Microlensing Observations in Astrophysics (MOA) collaboration and first alerted on 2020 July 7. This event was located at the J2000 equatorial coordinates $\left(\mathrm{R.A.,\ decl.}\right) = (17^\mathrm{h}\,53^\mathrm{m}\,41\overset{^\mathrm{s}}{.}64,\  -29\arcdeg48\arcmin\,27\overset{\prime\prime}{.}\,24)$, and Galactic coordinates $(l,\, b) =  (0.15598\arcdeg,\, -1.95678\arcdeg)$ in the MOA-II field `gb5' \citep{sumi2013microlensing}. The MOA observations were performed using the purpose-built 1.8m wide-field MOA telescope located at Mount John Observatory, New Zealand, and the observations of the field `gb5' were taken with a 15-minute cadence using the MOA-Red filter. The MOA-Red filter corresponds to a customized wide-band similar to a sum of the Kron-Cousins \textit{R} and \textit{I} bands, from $600\;\mathrm{nm}$ to $900\;\mathrm{nm}$. Occasional observations from the MOA group were made in the visual band using the MOA-\textit{V} filter.  The photometry in these filters was initially performed in real-time by the MOA pipeline \citep{bond2001real}, based on the difference imaging method \citep{tomaney1996expanding}. The data used in this paper are from a re-reduction done using the \citet{bond17} method, which performs a detrending process to correct for systematic errors and removes correlations in the data that may be present due to variations in the seeing and effects of differential refraction  \citep{bennett2012planetary, bond17}. The \citet{bond17} method also provides photometry calibrated to the Optical Gravitational Lensing Experiment phase III project (OGLE-III) \citep{szymanski2011optical}.

The Canada–France–Hawaii Telescope (CFHT), located near the summit of Mauna Kea  in Hawaii, United States, also observed the event in the SDSS\footnote{Sloan Digital Sky Survey \citep{fukugita1996sloan}} $i$-band filter.  From 2020 March to July, CFHT provided 1-2 supplementary observations per night toward the Korea Microlensing Telescope Network (KMTNet) high-cadence fields and follow-up observations for high-magnification events \citep{zang2021earth}. The CFHT data contributes to the establishment of the baseline brightness of the source  after the anomaly, and covers a two-day gap with a data point at $\mathrm{HJD}= 2,459,041.9$. The CFHT data were reduced by a custom difference imaging analysis pipeline \citep{zang2018measurement} based on the ISIS package \citep{alard1998method,alard2000image}.

The MOA-2020-BLG-135 event was also alerted one day later by the Korea Microlensing Telescope Network (KMTNet) group as KMT-2020-BLG-0579.  The KMTNet collaboration monitored this event with the 1.6m telescope located at the Siding Spring Observatory, Australia \citep{kim2016kmtnet}. Unfortunately, in addition to the KMTNet data not covering the anomaly, evidence of systematics was found in their data. Since the data would not improve the characterization of the main event and the planet, the KMTNet team suggested removing their data from the paper.

As a result of observatories' shutdowns due to the COVID-19 pandemic, data from KMT Cerro Tololo Interamerican Observatory, in Chile, KMT South African Astronomical Observatory, in South Africa, and OGLE, in Chile, could not be taken. These observatories were closed when the microlensing event happened.

Figure~\ref{fig:bestfit_model} shows the three datasets used for the analysis of this event. MOA-Red data and the MOA-\textit{V} data are displayed respectively in brown and violet colors, and the CFHT-\textit{i} data are in blue.

\section{Light-curve Models}\label{sec:lightcurve_models}
The light curve for the MOA-2020-BLG-135 event (see Figure \ref{fig:bestfit_model}) looks similar to a Paczy\'nski curve \citep{paczynski1986gravitational}, except for the anomaly in the interval $\mathrm{HJD'} = [9040.7, 9041.5]$\footnote{$\mathrm{HJD'} = \mathrm{HJD} - 2,450,000$} observed by both MOA-Red and MOA-\textit{V} (zoomed in Figure \ref{fig:bestfit_model}). The Paczy\'nski curve assumes a model in which the lens consists of a single star and the radiant flux comes from a single source. We display this curve as a point-source point-lens (PSPL) model in Figure \ref{fig:bestfit_model}. This deviation indicates that the lens may be composed of two masses, in which the less massive lens component can be a planet-mass object. We call this a binary or planetary lens system, depending on the mass ratio, as there are two objects contributing gravitationally as lenses. In Section~\ref{sec:lightcurve_2l1s}, we search for a lensing model explaining the three data sets presented in Section \ref{sec:data_photometry} (MOA-Red, MOA-\textit{V}, and CFHT-\textit{i}) by exploring the parameter space of possible binary lenses with a single source (2L1S) using the method described in \citet{bennett2010efficient}. In Section~\ref{sec:binarysource_exploration}, we discuss the lack of evidence for a binary-source model - with single lens (1L2S) or binary lens (2L2S) - after investigating the light curve using the method described in \citet{bennett2018first}.

\subsection{Single Source Scenario}
\label{sec:lightcurve_2l1s}
The \citet{bennett2010efficient} process uses the image-centered, ray-shooting method \citep{bennett1996detecting} combined with a custom version of the Metropolis algorithm \citep{metropolis1953equation}, which yields a rapid convergence to a $\chi^2$ minima. 

\subsubsection{Fit Parameters} \label{sec:lightcurve_parameters}
The parameters of our model are: the Einstein crossing time ($t_\mathrm{E}$); the time at which the separation of lens and source reaches the minimum ($t_0$); the minimum angular separation between source and lens as seen by the observer ($u_0$); the separation of the two masses of the binary lens system during the event ($s$); the counterclockwise angle between the lens-source relative motion projected onto the sky plane and the binary lens axis ($\alpha$); the mass ratio between the secondary lens and the primary lens ($q$); the source radius crossing time ($t_*$); the source flux for each instrument $i$ ($f_{s,i}$); and the blend flux per instrument $i$ ($f_{b,i}$).

The parameters $t_\mathrm{E}$, $t_0$ and $u_0$ are the common parameters for the single-lens model, while $s$, $\alpha$ and $q$ are the additional parameters for a binary lens system model. Both length parameters, $u_0$ and $s$, are normalized by the angular Einstein radius $\theta_\mathrm{E}$, defined by

\begin{equation}
\theta_\mathrm{E} = \sqrt{\frac{4GM_L}{c^2D_S}\left(\frac{D_S}{D_L}-1\right)},
\label{eqn:einstein_radius}
\end{equation}
where $G$ is the gravitational constant, $M_L$ is the mass of the lens system, $c$ is the speed of light, $D_S$ is the observer-source distance, and $D_L$ is the observer-lens distance. The source radius crossing time, $t_*$, is included in the lensing model to take account of finite source effects, 
\begin{equation}
t_* = \rho \: t_\mathrm{E} = \frac{\theta_*}{\theta_\mathrm{E}}t_\mathrm{E},
\label{eqn:crossing_time}
\end{equation}
where $\rho$ is the source angular radius in Einstein units, and $\theta_*$ is the source angular radius.

The other two parameters taken into account are the blend flux $f_{b,i}$ and the source flux $f_{s,i}$. As microlensing events are observed in crowded stellar fields, the source is usually blended with other unlensed stars. For this reason, we consider the blend flux. Since the observed brightness has a linear dependence on the blend flux and the source flux, they are treated differently from the other nonlinear fit parameters, as it follows. For every instrument and each set of the previously cited fit parameters, we can find a total flux that minimizes the $\chi^2$. The total flux $F_i(t)$ for time $t$ for instrument $i$ can be written as:
\begin{equation}
F_i(t) = A(t,\textbf{x})\:f_{s,i} + f_{b,i},
\label{eqn:flux}
\end{equation}
where $A(t,\textbf{x})$ is the magnification of the event at any given time and for any given set of nonlinear parameters $\textbf{x} = (t_\mathrm{E}, t_0, u_0, s, \alpha, q, t_*)$, $f_{s,i}$ is the unlensed source flux in a specific passband $i$, and $f_{b,i}$ is its excess flux. For many light curves reduced with difference imaging, the blend flux has an arbitrary normalization. Yet, the \citet{bond17} method normalizes the total flux to match the flux of the nearest star-like object in the reference frame, and it is the one used for the MOA data.

For this modeling, we do not consider parallax effects because this is a short ($t_\mathrm{E} \approx 17\; \mathrm{days}\ll 1\;\mathrm{month}$) and faint event. Additionally, its peak of magnification was reached on 2020 July 10, when the Earth’s instantaneous acceleration toward the projected position of the Sun projected into the lens plane was close to its minimum. These three factors make it very unlikely to detect asymmetric features in the light curve tails due to parallax effect. Therefore, we do not attempt a parallax measurement for this event.

\subsubsection{Exploring the Full Parameter Space} \label{sec:lightcurve_exploration}

{
\centering
\begin{table*}
\centering
\caption{Best-fit Model Parameters for MOA-2020-BLG-135, and Corresponding Medians from the Posterior Distribution}
\label{tab:model_parameters}
\begin{tabular}{lccccc}
\toprule
{Parameters}   &   {Units}         & {2L1S $s<1$} & {\textbf{2L1S} $\mathbf{s>1}$} &     {MCMC Medians} &         {$2\sigma$ Range} \\
\midrule
$t_\mathrm{E}$ &            {days} &     $16.905$ &       $16.791$ &          $16.85_{-0.27}^{+0.28}$ &          $16.30$-$17.41$ \\
$t_0$          &   $\mathrm{HJD'}$ & $9040.48797$ &   $9040.48843$ &   $9040.4879_{-0.0019}^{+0.0020}$ &  $9040.4839$-$9040.4920$ \\
$u_0$          &                   &    $0.03260$ &      $0.03287$ &  $0.03275 \pm 0.00072$ &     $0.03134$-$0.03421$ \\
$s$            &                   &    $0.95472$ &      $1.04509$ &        $0.997_{-0.054}^{+0.096}$ &         $0.916$-$1.126$ \\
$\alpha$       &         {radians} &    $2.38728$ &      $2.37731$ &        $2.383_{-0.013}^{+0.021}$ &          $2.360$-$2.433$ \\
$q$            &         $10^{-4}$ &    $1.55314$ &      $1.13963$ &           $1.52_{-0.31}^{+0.39}$ &           $1.01$-$2.47$ \\
$t_*$          &            {days} &    $0.14661$ &      $0.14186$ &        $0.145_{-0.011}^{+0.018}$ &         $0.125$-$0.183$ \\
$I_S$          &                   &     $19.009$ &       $19.000$ &       $19.004\pm 0.045$ &       $18.914$-$19.095$ \\
$V_S$          &                   &     $21.137$ &       $21.127$ &       $21.132\pm 0.045$ &       $21.042$-$21.223$ \\
fit $\chi^2$   &                   &    $14819.9$ &      $14819.7$ &                          \dots{} &                 \dots{} \\
\bottomrule
\end{tabular}
\end{table*}

}

We start modeling by systematically exploring the full parameter space with the grid-search approach described in \citet{bennett2010efficient}. For the first step, we do an initial condition grid search where $t_\mathrm{E}$, $t_0$, $u_0$, and $t_*$ are fixed, while $s$, $\alpha$, and $q$ are scanned. This is done so we can select the initial conditions. From there, we use the custom version of the Metropolis algorithm, as implemented by \citet{bennett2010efficient}, with initial positions coming from the 13 local solutions obtained from these initial scans. To ensure that no good model is missed in our analysis, we make sure that different mass ratios in the $q\in[10^{-5}, 10^{-1}]$ range are included in our set of initial conditions. Our customized Metropolis algorithm provides a full fit, in which all the parameters are free. Then, we select the best-fit model by looking for the lowest $\chi^2$, which indicates a model with $q \sim 10^{-4}$.

Interpreting a planetary lensing event with a planet signal is often subject to a close-wide degeneracy, in which a solution with a certain separation $s$ results in a similar model to another model presenting a solution with the same parameters' values but with a separation given by $1/s$.  This degeneracy arises from the symmetries in the lens equation \citep{griest1998use,dominik1999binary}. As discussed in \citet{yee2021ogle}, the degeneracy appears even for resonant caustics, which are far from the $s \gg 1$ limit in which the symmetries were derived. Therefore, we carefully cover both the close ($s<1$) and the wide ($s>1$) solutions in our analysis.

\begin{figure*}
  \includegraphics[width=0.99\linewidth]{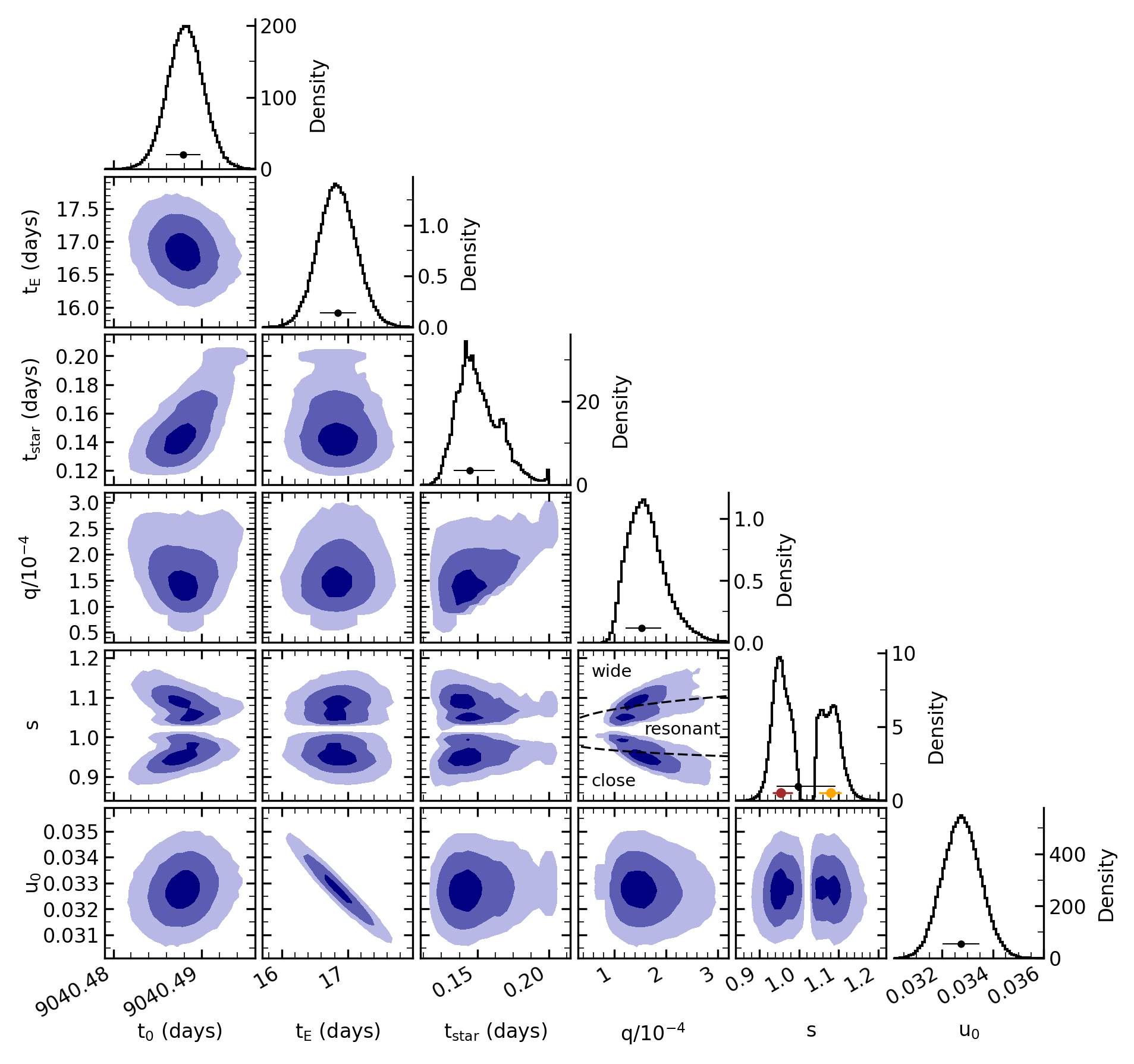}
  \caption{The marginalized posterior distributions for our MCMC runs, correlating the parameters for the binary lens single source model (2L1S), and, in the diagonal, the 1-dimension Probability Density Function (PDF) of each parameter. The $68.3\%$ (1$\sigma$), $95.5\%$ (2$\sigma$), and $99.7\%$ (3$\sigma$) confidence intervals are shown in dark blue, median blue, and light blue, respectively, in the posterior distribution plots. In the PDF plots, the black dot points out the median, and the thin line marks the 1$\sigma$ confidence interval. For the separation $s$ PDF plot, the additional red and yellow dots and lines also point out the median and the 1$\sigma$ confidence interval, but now for each of the two regions, the close-separation ($s<1$) in red and the wide-separation ($s>1$) in yellow. In panel $s$ vs. $q/10^{-4}$, the two dashed lines show the theoretical limits of the three caustic topologies (close, resonant, and wide).}
  \label{fig:mcmc_cornerplot}
\end{figure*}

\begin{figure}
  \subfloat[][Caustic geometry of the best-fit model for $s<1$, due to a close-separation planet.]{\includegraphics[width=\linewidth]{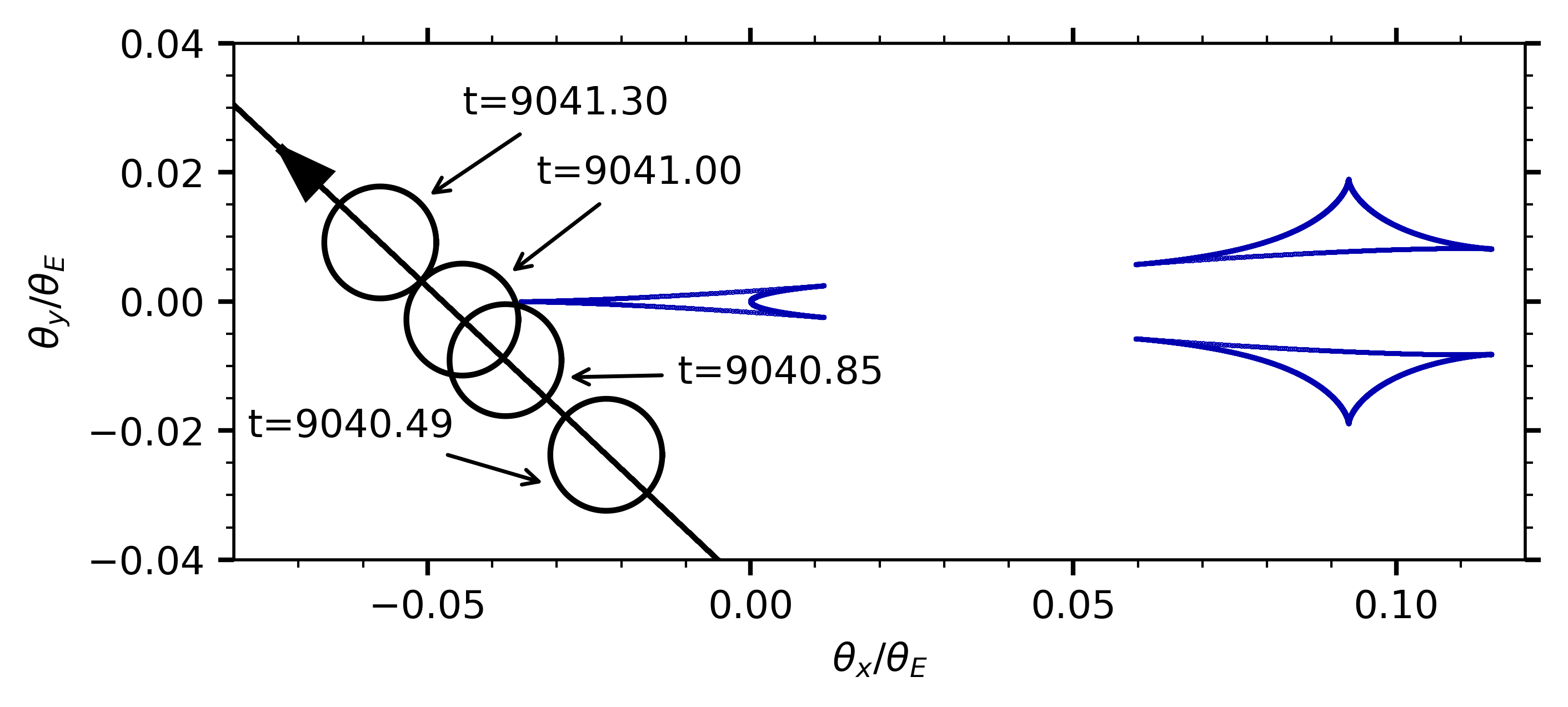} \label{fig:close_caustic}} \\
  
   \subfloat[][Caustic geometry of the best-fit model for $s>1$, due to a wide-separation planet.]{\includegraphics[width=\linewidth]{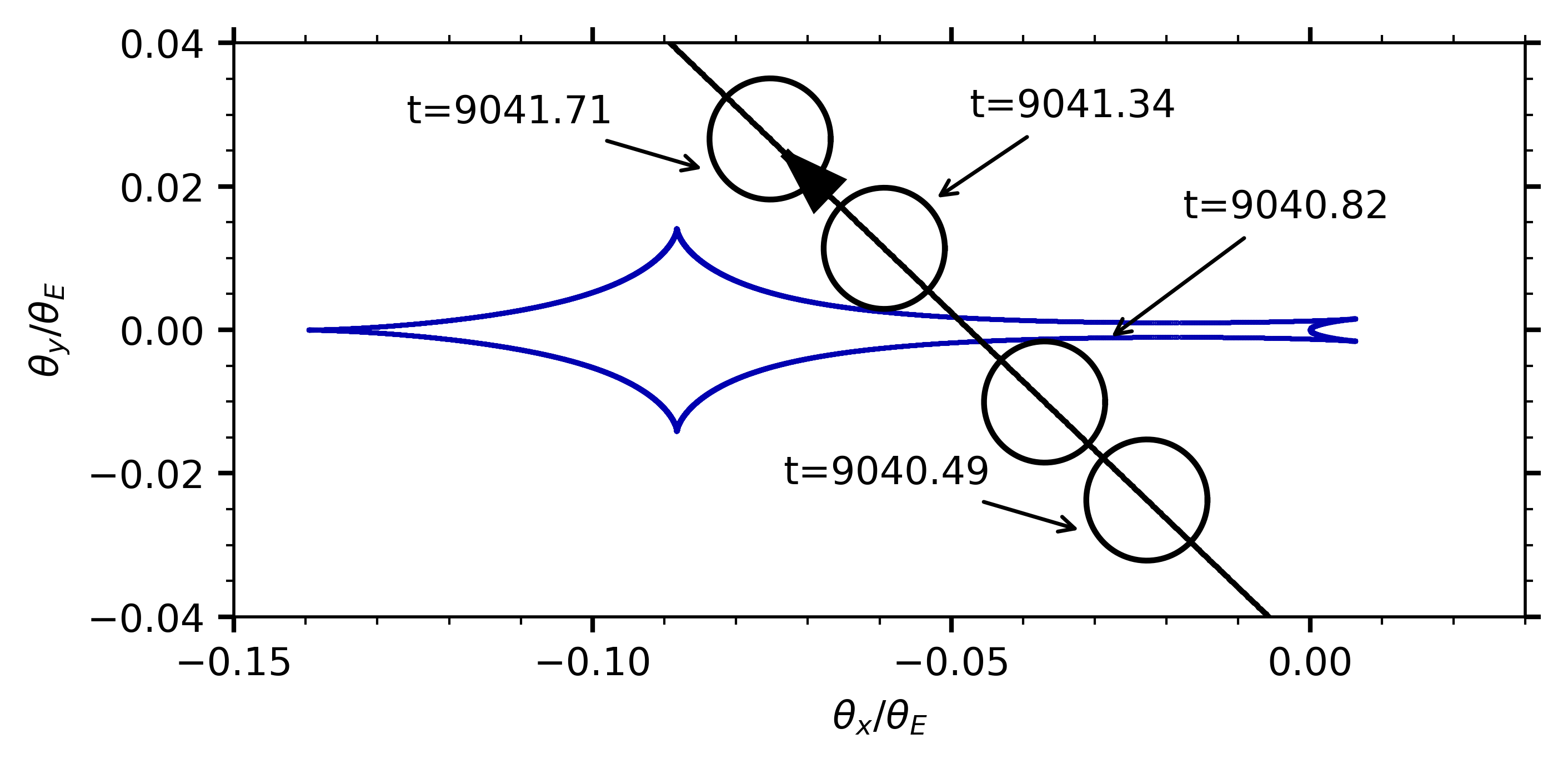} \label{fig:wide_caustic}}
   \caption{The caustic geometry and the source-lens trajectory. The caustic is represented in blue. The black straight solid line shows the source-lens trajectory and the arrow shows the direction of the source-lens relative motion. The source size is displayed as a black circle at its position at $\mathrm{HJD'} = 9040.49$ (event peak), $\mathrm{HJD'} = 9040.85$ for $s<1$ and $\mathrm{HJD'} = 9040.82$ for $s>1$(source starts crossing the caustic), and $\mathrm{HJD'} = 9041.00$ for $s<1$ and $\mathrm{HJD'} = 9041.34$ for $s>1$ (source exiting the caustic). For (b), the source at $\mathrm{HJD'} = 9041.71$ is also displayed to show when it is closest to the upper cusp.}
   \label{fig:caustics}
\end{figure}

To guarantee the exploration of the full parameter space with the Monte Carlo method, we use our customized version of the Metropolis algorithm for both the close and wide solutions. When running the Markov Chain Monte Carlo (MCMC) algorithm with the parameters of the best-fit models as initial inputs, we notice that the proposal distribution function we choose allows each chain to jump back and forth between the wide and close solutions. This happens because both values for the separation $s$ are close enough that the $\chi^2$-barrier between them has a size encompassed by the possible jumps. To ensure the optimization of the posterior sampling, we conduct MCMC runs adjusting the variables for our diagonalized covariance matrix and combine the results of the runs. For our analysis, we combine only the MCMC runs that jump between the wide and close solution, and vice-versa, and those have chains with a similar best-fit model.

Our best-fit planetary light curve model (2L1S $s>1$) is shown in Figure \ref{fig:bestfit_model}, and its parameters are given in Table~\ref{tab:model_parameters}, being the solution referred to as simply ``the best-fit model" in this paper. The result of our best-fit model for 2L1S $s<1$ is also displayed in Table~\ref{tab:model_parameters}. The median of the marginalized posterior distributions with the 1$\sigma$ confidence interval (i.e., $68.3\%$) is displayed in the same table, together with the range for distributions within the 2$\sigma$ interval (i.e., $95.5\%$). Figure \ref{fig:mcmc_cornerplot} shows the posterior distribution together with the 1$\sigma$, 2$\sigma$, and 3$\sigma$ (i.e., $99.7\%$) confidence intervals. 

In Figure~\ref{fig:mcmc_cornerplot}, a butterfly-wing shape is visible in the posterior distribution for $s$. This shape shows that both the close and wide solutions were fully explored during our MCMC runs. This is an effect of our algorithm jumping back and forth between both wide and close solutions. Even though there seems to exist two regions in each butterfly wing when considering the 1$\sigma$ interval, the chi-square difference is not big enough to create a barrier that could separate them into four independent regions. \citet{yang2022kmt} discuss the existence of pairs of close/wide solutions, which was named the ``central-resonant degeneracy". The mentioned interconnected four local minima in Figure ~\ref{fig:mcmc_cornerplot} could be interpreted as two of those pairs. Figure~\ref{fig:close_caustic} and Figure~\ref{fig:wide_caustic} illustrate the close-separation ($s<1$) and the wide-separation ($s>1$) topology, respectively. For the $s>1$ solution, a resonant caustic solution appears as the best solution, while for the $s<1$  solution, a non-resonant caustic solution is slightly better. It is not surprising that the $s>1$ and $s<1$ regions include resonant and non-resonant caustic solutions because the caustic topology is related to light curve features only in special cases. The $q/10^{-4}$ vs.\ $s$ panel of Fig~\ref{fig:mcmc_cornerplot} shows the dividing lines for these caustic topologies. For our data points, the best-fit model for the light curve with the planetary separation solution $s>1$ is practically the same as the one for the solution $s<1$. For $s<1$, the anomaly due to the presence of the planet appears to start at $\mathrm{HJD'} \approx 9040.85$, finishing at $\mathrm{HJD'} \approx 9041.30$, while for $s>1$, it appears to start at $\mathrm{HJD'} \approx 9040.82$, finishing at $\mathrm{HJD'} \approx 9041.34$. The explanation of these small differences can be found in Figure~\ref{fig:caustics}, which indicates the source starts and finishes crossing the caustic at slightly different times.  Moreover, we can check the great similarity for all the fitting parameters, together with both $s$ being close to the inverse of each other, indicating the historically called close-wide degeneracy solutions. 

Examples of close-wide model degeneracies which do not obey the $s\leftrightarrow 1/s$ relationship predicted by \citet{dominik1999binary} are relatively common \citep[e.g.][]{bennett2014moa,koshimoto-ob120950}, and \citet{an2021condition} pointed out that many such cases can be explained similarities of local caustic regions even in situations in which the overall caustic shape may not be degenerate at all. Adding to the degeneracy discussion, \citet{zhang2022ubiquitous} have recently demonstrated that the close-wide degeneracy $s\leftrightarrow 1/s$ relationship is only strictly followed for the singular case of $u_0=0$. For the general case of $u_0>0$, the authors proposed an alternative theory named ``the offset degeneracy" that predicts a deviation from $s\leftrightarrow 1/s$. The formalism is shown to be mathematically exact in certain limits \citep{zhang2022mathematical}, which includes caustic-crossing events and resonant events. In this work, the different degenerate solutions do not indicate distinct conclusions for the posterior distribution of the physical properties. Therefore, the solutions have a similar lens physical interpretation.

One might wonder about the proximity of the source and the upper cusp at $\mathrm{HJD'} \approx 9041.71$ in the lower panel of Figure~\ref{fig:caustics}. This upper cusp does not seem to create an extra perturbation, and it is located far from the source, more than one source radius away at $\mathrm{HJD'} \approx 9041.71$. It results in no apparent anomaly in the best-fit model in Figure~\ref{fig:bestfit_model}.

\subsection{Binary Source Possibility}
\label{sec:binarysource_exploration}

The investigation of the possibility of two sources was encouraged when still considering using the KMTNet data. However, the evidence of binary source largely disappeared when systematic of KMTNet data was discovered.

{
\centering
\begin{table}
\centering
\caption{Comparison Between Microlensing Models}
\label{tab:comparison_model}
\begin{tabular}{ccrr}
\toprule
Model                          &   $N_\mathrm{parameters}$ &   $\Delta\chi^2$ & $\Delta\mathrm{BIC}$\\
\midrule
PSPL/1L1S                      &                         3 &           669.28 & 630.87 \\
1L2S                           &                         8 &            42.53 &  52.13 \\
2L1S s\;$ < $\;1               &                         7 &             0.21 &   0.21 \\
\textbf{2L1S} $\mathbf{s>1}$   &                         7 &          \dots{} & \dots{}\\
2L2S                           &                        12 &           -12.17 &  35.85 \\

\bottomrule
\end{tabular}
\end{table}
}

Following the suggestion of the KMTNet collaboration, we remove the KMTNet data from the analysis, as explained in Section~\ref{sec:data_photometry}, and we attempt to fit binary source models to our data. We use a similar method to the one described in \citet{bennett2018first} to look for solutions with binary sources. First, we search for solutions within binary-source single-lens (1L2S) models and, then, within binary-source binary-lens (2L2S) models. The 1L2S models are obtained with the same code used for the 2L2S models, but with the mass of one of the lenses set to 0. For easier comparison, we refer to the Point-Source Point-Lens (PSPL) model as PSPL/1L1S, the best-fit model as 2L1S $s>1$, and its degenerate model as 2L1S $s<1$ (Section~\ref{sec:lightcurve_2l1s}).

\begin{figure}
  \includegraphics[width=\linewidth]{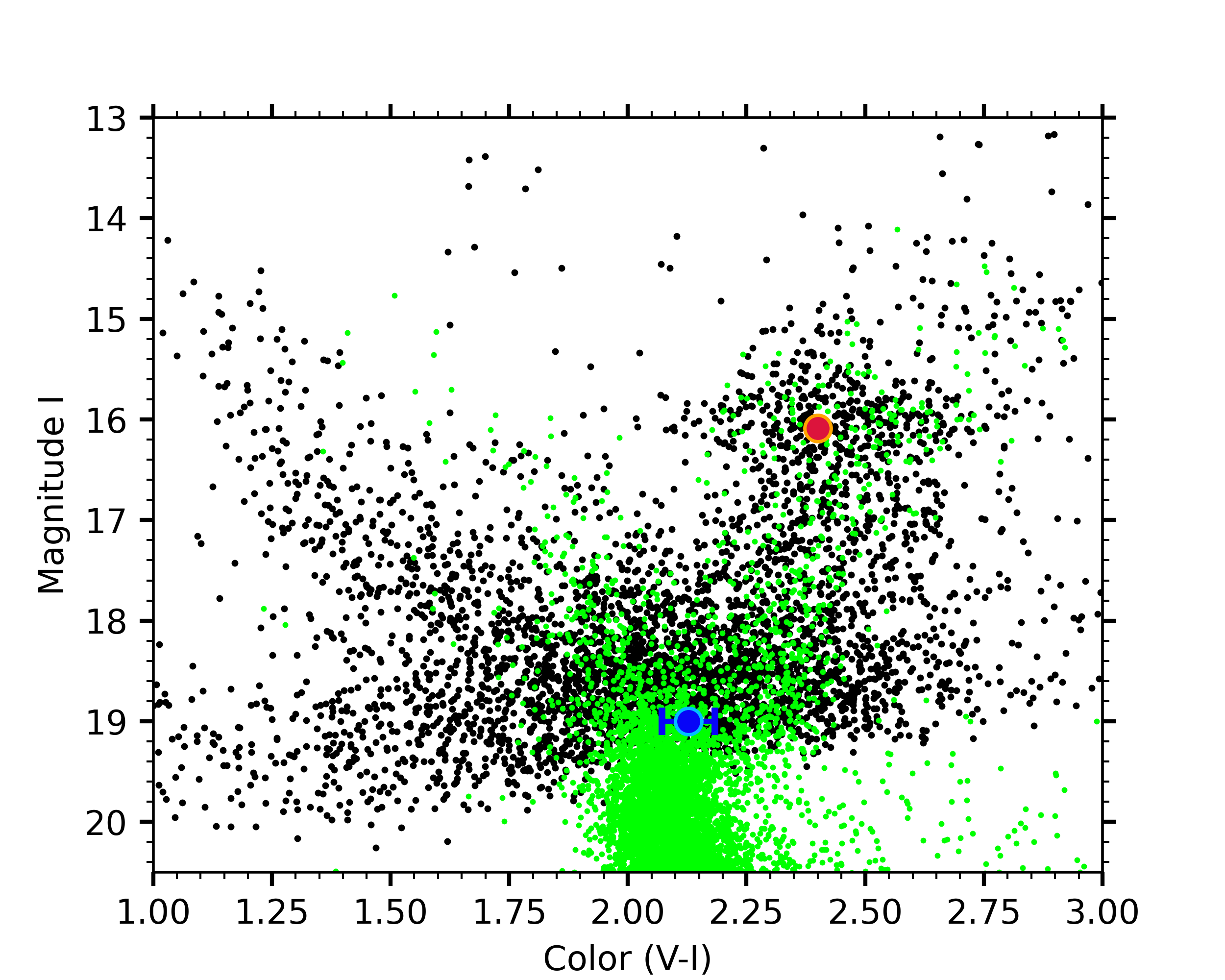}
  \caption{(V - I, I) Color-magnitude diagram of the stars in the OGLE-III catalog within $90^{\prime\prime}$ of MOA-2020-BLG-135. The black dots are the stars from the OGLE-III catalog, the blue dot indicates the source magnitude and color for the best-fit model ($s>1$), and the red circle indicates the red clump giant centroid. For comparison, we added the green dots showing the Hubble Space Telescope CMD from \citet{holtzman1998luminosity} shifted to the bulge distance and relevant extinction derived in Section~\ref{sec:photometric_calibration_source_radius}. The source star is probably in a subgiant phase.}
  \label{fig:cmd}
\end{figure}

The \citet{bennett2018first} method allows us to add a second source with different brightness and color from the first source. We find the best-fit for the 1L2S model to have a fit $\chi^2 = 14862.24 $, with a lens-source proper motion of $\mu_{\mathrm{rel,G}} = 0.44 \pm 0.05 \mathrm{\;mas\;yr^{-1}}$, which is unusually small. With $\Delta \chi_\mathrm{1L2S-2L1S}^2 = 42.53 $ and a small relative proper motion, our results indicate a strong preference toward the 2L1S $s>1$ model. Therefore, the 1L2S model is ruled out because it suggests that a model with a second source and only one lens reduces the quality of the fitting to the data. 

For the binary-source binary-lens, 2L2S, model, we include the binary-lens parameters. We obtain a fit $\chi^2 =14807.54$, which is smaller than the one for the 2L1S  $s>1$ model ($\Delta \chi_\mathrm{2L1S-2L2S}^2 = -12.17$). As expected, due to a larger number of parameters, a model with 2L2S may improve the $\chi^2$, so an extra criterion is necessary to evaluate the significance of this small improvement. Therefore, we also calculate the Bayesian Information Criterion (BIC) \citep{schwarz1978estimating} indexes for the two models. The difference in BIC index is $\Delta\mathrm{BIC}_\mathrm{2L2S - 2L1S}=35.85$. The improvement of the quality of the fit is, then,  proved to be insignificant when using the BIC index. Moreover, it should be noted that a binary source event requires special alignment of the sources, being generally unlikely to be detected. Therefore, we continue our analysis considering only the 2L1S model. 

The differences in $\chi^2$ and BIC index for each model compared to our best-fit model is displayed in Table~\ref{tab:comparison_model}.

\section{Photometric Calibration and Source Properties}\label{sec:photometric_calibration_source_radius}

The source angular radius $\theta_*$ is not explicitly obtained from the lightcurve models presented in Section \ref{sec:lightcurve_models}, yet it can be empirically derived if we know the de-reddened magnitude and color of the source \citep{van1999predicting, yoo2004ogle,bennett2010efficient}. Toward this aim, we correct the extinction and reddening for the source star by using the red clump giants as a reference.

{
\centering
\begin{table}
\centering
\caption{Source and lens-source properties}
\label{tab:source_properties_mcmc}
\begin{tabular}{lcr}
\toprule
Parameters &    Units &               MCMC Medians \\
\midrule
Source magnitude $I_\mathrm{S,0}$                &                          &  $17.350_{-0.078}^{+0.078}$ \\
Source magnitude $K_\mathrm{S,0}$                &                          &    $16.45_{-0.25}^{+0.22}$ \\
Source color $(V-I)_\mathrm{S,0}$                &                          &  $0.788_{-0.096}^{+0.095}$ \\
Source angular radius $\theta_*$                 &        $\mathrm{\mu as}$ &     $1.15_{-0.12}^{+0.13}$ \\
Einstein radius $\theta_\mathrm{E}$              &           $\mathrm{mas}$ &  $0.133_{-0.018}^{+0.019}$ \\
Lens-source proper motion $\mu_{\mathrm{rel,G}}$ &  $\mathrm{mas\;yr^{-1}}$ &      $2.88_{-0.40}^{+0.42}$ \\
\bottomrule
\end{tabular}
\end{table}
}

In order to obtain the dereddened color and corrected magnitude of our source star, we first calibrate the instrumental MOA-II magnitudes, MOA-Red and MOA-\textit{V}, with the OGLE-III catalog considering the following equations from the standard calibration procedure described in \citet{bond17}:

\begin{multline}
    I_\mathrm{O3} = (28.0983 \pm 0.0014) + R_{MOA} \\ - (0.20844 \pm 0.00087) \times (V_{MOA} - R_{MOA}) \label{eq:i_band} \\
 \end{multline}
 \begin{multline}  
    V_\mathrm{O3} = (28.5038 \pm 0.0014) + V_{MOA} \\ - (0.10746 \pm 0.00088) \times (V_{MOA} - R_{MOA}) \label{eq:v_band},
 \end{multline}
where $R_\mathrm{MOA}$ is the MOA-Red filter and $V_\mathrm{MOA}$ is the MOA-\textit{V} filter. The \citet{bond17} calibration is the standard MOA calibration procedure and is done using cross-matched stars within the $500\arcsec  \times 500\arcsec$ cameo image from our DOPHOT \citep{schechter1993dophot} catalog  with stars from the OGLE-III catalog. In the final fit, the root mean square scatters, which are dominated by thousands of faint stars, are $\mathrm{RMS}_I \approx 0.07$ and $\mathrm{RMS}_V \approx 0.08$. We then obtain the magnitude $I_\mathrm{O3}$ and $V_\mathrm{O3}$ \citep{gould2010second} for our source star. These are displayed in Table \ref{tab:model_parameters} as $I_\mathrm{S}$ and $V_\mathrm{S}$ respectively. Figure \ref{fig:cmd} shows the calibrated color and magnitude for our source star, compared to the stars within a $90^{\prime\prime}$ radius limit from our target.  

For the second step, we measure the extinction and reddening of the stars within a $90^{\prime\prime}$ radius limit from our source star as follows. First, we calculate what are the apparent magnitude and color of the red clump, obtaining $I_\mathrm{RCG} = (16.09 \pm 0.05)$ and $(V-I)_\mathrm{RCG} = (2.40 \pm 0.05)$ (represented as the red dot in Figure \ref{fig:cmd}). The expected dereddened magnitude of the red clump at a Galactic longitude $l=0.15598\arcdeg$ is $I_\mathrm{RCG,0} = (14.44 \pm 0.04)$~\citep{nataf2013reddening}, and the expected color is $(V-I)_\mathrm{RCG,0} = (1.06 \pm  0.06)$ \citep{bensby2011chemical}. Therefore, we can use the calculated apparent magnitude and color of the red clump in comparison with the expected values to obtain the extinction and the color excess of this region in the sky. The extinction and the color excess are: $A_\mathrm{I} = (1.65 \pm 0.06)$ and $E(V-I) = (1.34 \pm 0.08)$. These values are reasonable, and can be compared to the ones given by the OGLE-III tool for querying interstellar extinction toward the Galactic bulge \footnote{http://ogle.astrouw.edu.pl/cgi-ogle/getext.py, based on \citet{nataf2013reddening}.}, which were $A_\mathrm{I} = 1.60$ and $E(V-I) = 1.35$ when using the natural neighbor interpolation option. 

With our calculated extinction and the color excess, we obtain the corrected magnitude  $I_\mathrm{S,0} = (17.350 \pm 0.078)$ and dereddened color $(V-I)_\mathrm{S,0} = 0.788_{-0.096}^{+0.095}$ (in Table \ref{tab:source_properties_mcmc}) of our source star.

Finally, we determine the source size:
\begin{equation}
\log_{10}\left[\dfrac{2\theta_*}{\mathrm{mas}}\right] = 0.501414\;+\; 0.419685 (V-I)_\mathrm{S,0}\;-\;0.2 I_\mathrm{S,0}
\label{eqn:boyajian}
\end{equation}
following \citet{boyajian2014stellar} analysis for stars with $3900\, \mathrm{K}<T_\mathrm{eff}<7000\,\mathrm{K}$, and appearing in \citet{bennett2017moa}. The source angular radius is $\theta_* = 1.15_{-0.12}^{+0.13}$ $\mu as$.

Determining the source size is necessary to calculate the angular Einstein radius $\theta_\mathrm{E}$ when using Equation \ref{eqn:crossing_time}. The importance of $\theta_\mathrm{E}$ itself comes from the fact that the main physical properties of any microlensing event (the lens mass $M_\mathrm{L}$, the distance to the lens $D_\mathrm{L}$, the distance to the source $D_\mathrm{S}$, and the lens-source relative proper motion $\mu_{\mathrm{rel,G}}$) are directly related to it. The measurement of $\theta_\mathrm{E}$ provides one mass-distance relationship, defined in Equation~\ref{eqn:einstein_radius}, which can be rearranged into

\begin{equation}
M_\mathrm{L} = 0.1228 M_\odot \left(\frac{\theta_\mathrm{E}}{1 \;\mathrm{mas}}\right)^2  \left(\frac{\pi_\mathrm{rel}}{1 \;\mathrm{mas}}\right)^{-1}, \mathrm{and}
\label{eqn:mass_einstein}
\end{equation}

\begin{equation}
D_\mathrm{L} = 1\;\mathrm{kpc} \left(\frac{\pi_\mathrm{rel}}{1 \;\mathrm{mas}} + \left(\frac{D_\mathrm{S}}{1 \;\mathrm{kpc}}\right)^{-1}\right)^{-1},
\label{eqn:distance_einstein}
\end{equation}
where the lens-source relative parallax $\pi_\mathrm{rel}$ is defined as

\begin{equation}
\pi_\mathrm{rel} = \frac{1\;\mathrm{AU}}{D_\mathrm{L}} - \frac{1\;\mathrm{AU}}{D_\mathrm{S}}.
\label{eqn:lenssource_parallax}
\end{equation} 

We use the Einstein crossing time $t_\mathrm{E}$ and the source radius crossing time $t_*$ from our light curve model, combined with our calculated source angular radius $\theta_*$, to obtain the angular Einstein radius $\theta_\mathrm{E} = 0.133_{-0.018}^{+0.019} \;\mathrm{mas}$. Even though we can define the lens-source proper motion $\mu_{\mathrm{rel,G}}$ as a function of $\theta_\mathrm{E}$, we use the following formula for $\theta_*$ and $t_*$ that avoids an increased uncertainty due to the blending degeneracy: 

\begin{equation}
\mu_{\mathrm{rel,G}} = \frac{\theta_\mathrm{E}}{t_\mathrm{E}} = \frac{\theta_*}{t_*}.
\label{eqn:lens_proper_motion}
\end{equation}
The lens-source proper motion is $\mu_{\mathrm{rel,G}}= 2.88_{-0.40}^{+0.42} \mathrm{\;mas\;yr^{-1}}$.

Aiming to provide more information about the source for future high-angular resolution follow-up observations, we also estimate the magnitude of the source in $K$-band by using the color transformations presented in \citet{kenyon1995pre}. The source magnitude is $K_\mathrm{S,0} = 16.45_{-0.25}^{+0.22}$ without extinction. By adding the extinction $A_K = 0.2195$ \citep{nishiyama2009interstellar,gonzalez2012reddening}, we obtain $K_\mathrm{S}=16.67_{-0.25}^{+0.22}$ as the magnitude of the source in $K$-band.

The calculated source and lens-source properties described in this section are in Table \ref{tab:source_properties_mcmc}. 

\section{Physical Properties of the Lens}\label{sec:lens_system_properties}

{
\centering
\begin{table*}
\centering
\caption{Lens physical properties derived from the \citep{bennett2014moa} galactic model.}
\label{tab:lens_properties}
\begin{tabular}{lcp{3.0cm}p{1.5cm}p{3.5cm}p{1.5cm}}
\toprule
{} Parameters &       Units & Prior uniform in M, $ \diff P \propto \diff M $  & $2\sigma$ Range ($95.5\%$)& Prior proportional to M, $ \diff P \propto M \diff M$ & $2\sigma$ Range ($95.5\%$) \\
\midrule
Planet mass $m_\mathrm{planet}$        &  $M_\oplus$ &            $11.3_{-6.9}^{+19.2}$ &    $2.1$-$57.3$ &                $25_{-15}^{+22}$ &    $4$-$70$ \\
Host mass $M_\mathrm{host}$            &   $M_\odot$ &           $0.23_{-0.14}^{+0.39}$ &   $0.05$-$1.07$ &                $0.53_{-0.32}^{+0.42}$ &   $0.08$-$1.22$ \\
Lens distance $D_\mathrm{L}$           &       $kpc$ &           $7.9_{-1.0}^{+1.0}$ &    $5.9$-$9.7$ &                 $8.3_{-1.0}^{+0.9}$ &    $6.4$-$9.9$ \\
Projected separation $a_\perp$         &        $AU$ &            $1.11_{-0.20}^{+0.23}$ &   $0.74$-$1.58$ &                 $1.17_{-0.20}^{+0.23}$ &    $0.80$-$1.64$ \\
Deprojected separation $a_\mathrm{3D}$ &        $AU$ &           $1.35_{-0.32}^{+0.75}$ &   $0.82$-$4.72$ &                $1.42_{-0.32}^{+0.78}$ &   $0.88$-$4.97$ \\
Lens magnitude $I_\mathrm{L}$          &             &             $26.0_{-2.9}^{+2.4}$ &   $20.2$-$36.7$ &                  $23.8_{-3.1}^{+2.5}$ &   $19.5$-$28.7$ \\
Lens magnitude $K_\mathrm{L}$          &             &             $22.2_{-2.3}^{+1.9}$ &   $17.6$-$32.0$ &                  $20.4_{-2.3}^{+2.0}$ &   $17.1$-$24.2$ \\
\bottomrule
\end{tabular}
\end{table*}
}

As parallax and lens brightness measurements are missing for the MOA-2020-BLG-135 event, we cannot uniquely determine the lens mass and its distance. Therefore, to estimate the lens properties, we use the \citet{bennett2014moa} Galactic model. The strength of this model is that it can incorporate a prior for the Bayesian analysis when estimating the posterior probability distribution of the host mass. It allows us to use a mass function under the most conventional assumption that all stars have an equal planet-hosting probability, or under the assumption that planets are more likely to orbit around more massive stars, by setting a prior proportional to $M$.

\begin{figure*}
  \includegraphics[width=\linewidth]{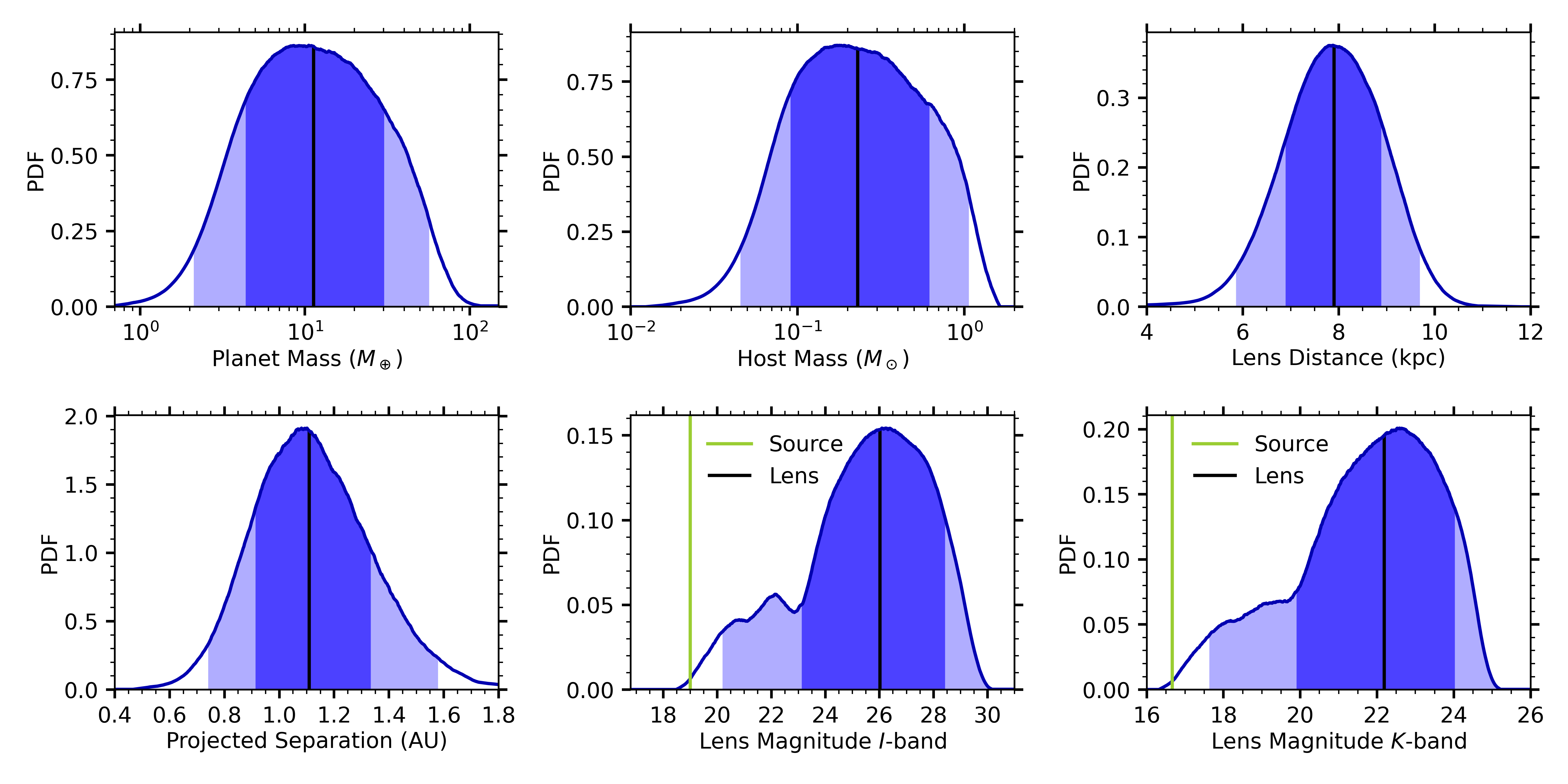}
  \caption{Lens system properties derived from the \citet{bennett2014moa} Galactic model using a power-law stellar mass function with a prior uniform in $M$, $\diff P \propto \diff M$, meaning all stars have an equal planet-hosting probability. The  1$\sigma$ and 2$\sigma$ confidence intervals (i.e., $68.3\%$ and $95.5\%$) are represented by dark blue and median blue,   respectively, with the median marked as a solid black line. For comparison, the source magnitude is shown as a green line in the lens magnitude graphs.}
  \label{fig:galactic_model_0}
\end{figure*}

Usually, in microlensing papers, the estimations for the lens system properties assume that all the stars have equal probability of hosting a planet, which implies a mass function prior uniform in $M$. Statistical results on exoplanet populations were also obtained under the same assumption \citep{Cassan2012}. Yet, for this paper, we also consider a second scenario in which the probability of hosting a planet scales in proportion to the host star mass, $\diff P \propto M\diff M$. \citet{johnson2007new,johnson2010giant} found that, for their radial velocity sample, the planet occurrence increases with the stellar mass at fixed planet mass. This is compatible with \citet{nielsen2019gemini} direct imaging sample, which showed a strong correlation between planet occurrence rate and host star mass. Moreover, \citet{bhattacharya2021moa} identified that the traditional assumption, which considers that all the stars have equal probability of hosting a planet, is not consistent with many microlensing events that have been revisited with the help of the Keck adaptive optics and had their lens object identified using high-angular resolution observations. \citet{bhattacharya2021moa} pointed out that five of the six events with direct measurement of the separation between the source and the lens stars have found a host star more massive than the median predicted under the most conventional assumption, which is the one with a prior uniform in $M$. The authors also indicated that there is no publication bias for that Keck sample. Certainly, a more extensive sample to state this with more confidence is needed and, in fact, NASA Keck Key Strategic Mission Support and \textit{Hubble Space Telescope} observing programs will be directly measuring the mass of more microlens host stars. Although planets are more likely to orbit around more massive stars, we still decide to consider both mass priors for our Bayesian assumptions, the conventional prior uniform in the stellar mass, and a prior that scales in proportion to the stellar mass.

\begin{figure*}
  \includegraphics[width=\linewidth]{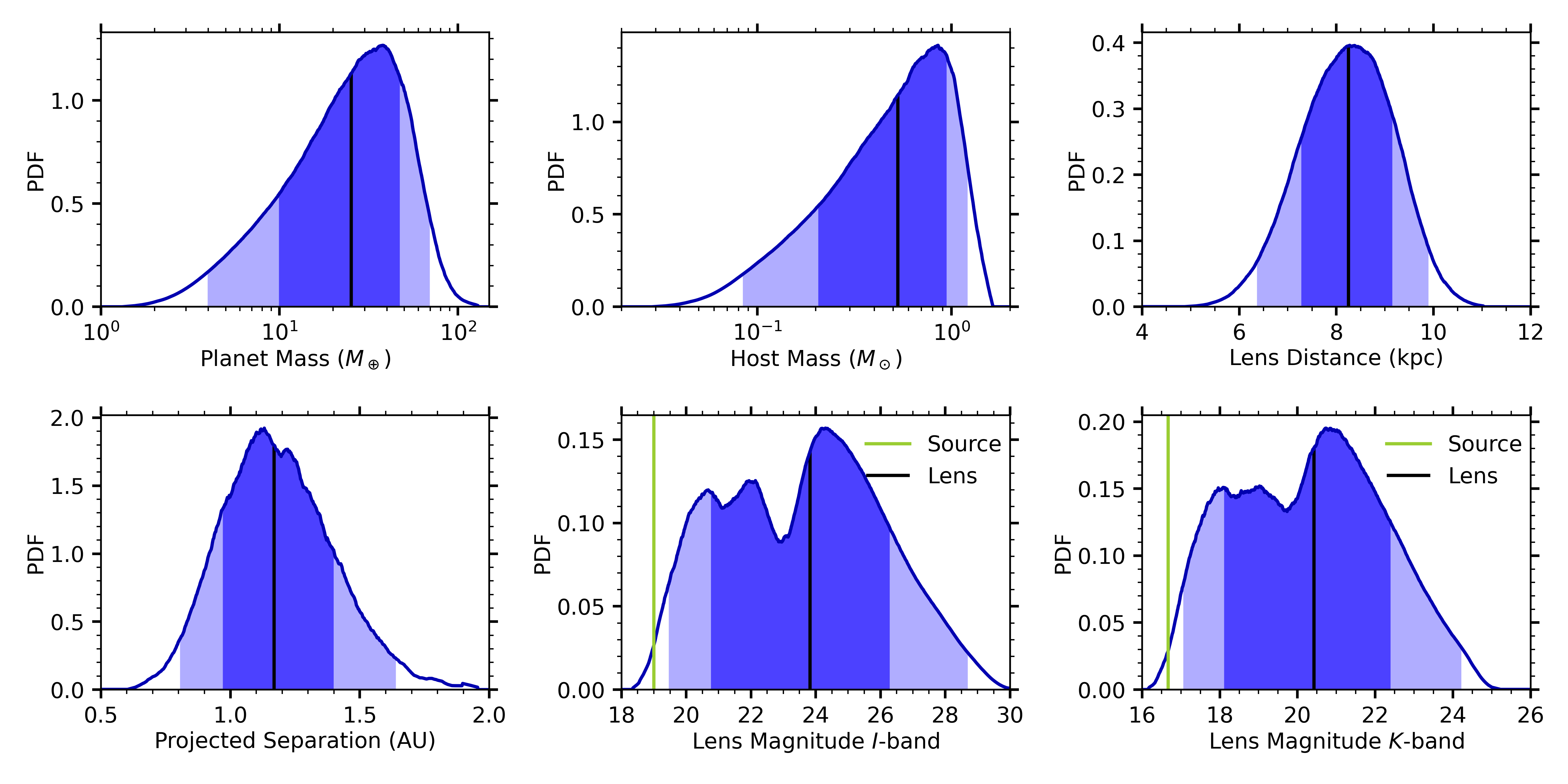}
  \caption{Lens system properties derived from the \citet{bennett2014moa} Galactic model using a power-law stellar mass function under the assumption that the hosting probability scales in proportion to the stellar mass, $\diff P \propto M \diff M$. The 1$\sigma$ and 2$\sigma$ confidence intervals (i.e., $68.3\%$ and $95.5\%$) are represented by dark blue and median blue, respectively, with the median marked as a solid black line. For comparison, the source magnitude is shown as a green line in the lens magnitude graphs.}
  \label{fig:galactic_model_1}
\end{figure*}

Results of the Bayesian analysis can also be affected when the probability of having a planet depends on the stellar location in our galaxy. However, it has recently been shown by \citet{koshimoto2021no} that the dependence of the planet-hosting probability on the Galactocentric distance is not large, thus we do not consider such dependence in this paper.

The Bayesian analysis for the lens properties is done using as input the collection of the MCMC runs (discussed in Section \ref{sec:lightcurve_exploration}), with their fit parameters, along with our calculated angular source radius and extinction (see Section \ref{sec:photometric_calibration_source_radius}). To obtain the magnitude of the lens not only in the $I$-band, but also in the $K$-band, which is useful for future high-angular resolution follow-up observations, we use the extinction $A_K = 0.2195$ \citep{gonzalez2012reddening, nishiyama2009interstellar}. The \citet{bennett2014moa} Galactic model is used two times, with the two different priors. First, we consider a power-law stellar mass function under the standard assumption that all stars have an equal planet-hosting probability, a function of the form $\diff P \propto M^\beta \diff M$ with $\beta = 0$. Then we consider the same function but under the assumption that planets are more likely to orbit around more massive stars, so a function with $\beta=1$.

Under the assumption that the probability of hosting a planet is the same for all stars, $\beta = 0\Rightarrow \diff P \propto \diff M$, we estimate that the lens physical properties and their $1\sigma$ ($68.3\%$) interval of confidence are $m_\mathrm{planet} = 11.3_{-6.9}^{+19.2} \;M_\oplus$ for the planet mass, $M_\mathrm{host} = 0.23_{-0.14}^{+0.39} \;M_\odot$ for the host mass, $D_\mathrm{L} = 7.9_{-1.0}^{+1.0} \;\mathrm{kpc}$ for the distance to the lens, $a_\perp = 1.11_{-0.20}^{+0.23} \;\mathrm{AU}$ for the project separation, $a_\mathrm{3D} = 1.35_{-0.32}^{+0.75} \;\mathrm{AU}$ for the deprojected separation, and $I_\mathrm{L} = 26.0_{-2.9}^{+2.4}$ and $K_\mathrm{L} = 22.2_{-2.3}^{+1.9}$ for the lens magnitude. When assuming the mass function prior proportional to $M$, $\beta = 1 \Rightarrow \diff P \propto M\diff M$, the planet mass estimation is $m_\mathrm{planet} = 25_{-15}^{+22} \;M_\oplus$, the host mass is $M_\mathrm{host} = 0.53_{-0.32}^{+0.42} \;M_\odot$, the distance to the lens is $D_\mathrm{L} = 8.3_{-1.0}^{+0.9} \;\mathrm{kpc}$, the projected separation is $a_\perp = 1.17_{-0.20}^{+0.23} \;\mathrm{AU}$, the deprojected separation is $a_\mathrm{3D} = 1.42_{-0.32}^{+0.78} \;\mathrm{AU}$, and lens magnitude is $I_\mathrm{L} = 23.8_{-3.1}^{+2.5}$  and $K_\mathrm{L} = 20.4_{-2.3}^{+2.0}$.  We report the probability distributions considering both priors in Table \ref{tab:lens_properties}.

Figure \ref{fig:galactic_model_0} shows the probability distribution of the planet and host masses, the distance to the lens system, their projected separation, and the lens magnitudes in both $I$-band and $K$-band, under the assumption of equal planet-hosting probability, $ \diff P \propto \diff M $. Figure \ref{fig:galactic_model_1} shows the same results, but for the probability scaling in proportion to the host mass, $ \diff P \propto M \diff M $. It is interesting to note that, even though the host mass and the planetary mass medians  seem to be almost the double when comparing the results when the prior is uniform in $M$ to the results when the prior is proportional to $M$, the range for the masses are not that different when considering the $2\sigma$ (i.e., $95.5\%$) confidence interval.

\section{Discussion}\label{sec:discussion}
Our light-curve analysis for the event MOA-2020-BLG-135 leads to the discovery of MOA-2020-BLG-135Lb, a new Neptune-class planet. This analysis yields a planet-host mass ratio of $q = 1.52_{-0.31}^{+0.39} \times 10^{-4}$, and separation $s\approx1$. It is important to mention that with each MCMC chain we were able to sample all the modes of the posterior distribution due to the closeness of our close-separation ($s<1$) and wide-separation ($s>1$) solutions.

In Sections \ref{sec:photometric_calibration_source_radius} and \ref{sec:lens_system_properties}, we determine the source and the lens magnitude in \textit{K}-band, aiming to anticipate results for high-angular resolution follow-up observations. The source magnitude with added extinction was computed to be $K_\mathrm{S}=16.67_{-0.25}^{+0.22}$.
Under the assumption that all stars have equal planet-hosting probability, the lens magnitude is expected to be in the range $19.9-24.1$ for the $68.3\%$ ($1\sigma$) confidence interval, and $17.6-32.0$ for the $95.5\%$ ($2\sigma$) interval, with median $22.2$. One might wonder whether the lens star looks faint in comparison with the source star when considering only the median and the $1\sigma$ interval, which can indicate that its observation might be challenging in near future. However, for this assumption, when considering the $2\sigma$ interval of confidence, the result looks more encouraging. The event MOA-2007-BLG-400 had its lens object successfully identified using high-angular resolution observations, and the magnitude for the source and lens in Keck $K$-band were, respectively, $16.43 \pm 0.04$ and $18.93 \pm 0.08$ \citep{bhattacharya2021moa}, which are not that different from our event, when considering the brightest possible lens magnitude. Moreover, on the assumption that the probability of hosting a planet scales in proportion to the stellar mass, the scenario gets even better, the lens magnitude is expected to be in the range $18.1-22.4$ for the $68.3\%$ ($1\sigma$) confidence interval, and $17.1-24.2$ for the $95.5\%$ ($2\sigma$) interval, with median $20.4$. Knowing that the microlensing events tend to have a host star more massive than their median predictions from only the light curve analysis, direct detection of the lens star looks promising for our event MOA-2020-BLG-135.

Another notable detail from our results is that a planet with a mass ratio $q = 1.52_{-0.31}^{+0.39} \times 10^{-4}$ seems to be exactly where previous core accretion theories predicted a Neptune desert \citep{ida2004toward,mordasini2009extrasolar}. It also lies exactly in the peak of the planet-to-star mass ratio distribution measured by the state-of-the-art statistical analysis of planets detected by gravitational microlensing \citep{suzuki2016exoplanet}. Recently, \citet{zang2022systematic} reported the analysis of a statistical planetary sample with 13 homogeneously-selected planets observed in 2019 by the KMTNet Collaboration. The paper suggests that the mass ratio function may not decrease rapidly below the \citet{suzuki2016exoplanet} mass-ratio break ($q_{br} \sim 10^{-4}$), instead, there may be a uniform distribution in $\log q$. Once more, the \citet{suzuki2016exoplanet} contained 30 planets and \citet{zang2022systematic} contained 13 planets. To better understand this planet-to-star mass ratio distribution, we need a larger microlens exoplanet sample. The MOA collaboration has been working to obtain this extended sample, and will have more than 50 new planets, including the planet presented in this paper. Meanwhile, the KMTNet collaboration is also working on a sample, predicting about 120 planets from 2016 to 2019. It is important to notice that the specifications of the telescopes and their cadence and sensitivity are different, therefore, both analysis are necessary.


In Section~\ref{sec:lens_system_properties}, we discussed the Bayesian priors used for the analysis of the lens system properties subject to the microlensing light curve constraints. In almost all previous analyses with similar light curve constraints, it has been assumed that all host stars have an equal probability to host the planet with the measured mass ratio ($q = 1.52_{-0.31}^{+0.39} \times 10^{-4}$ for this event). However, \citet{bhattacharya2021moa} have shown that higher mass host stars appear to be more likely for the planets that the microlensing method is sensitive to. This is somewhat similar to earlier findings from radial velocity \citep{johnson2007new,johnson2010giant} and direct detection \citep{nielsen2019gemini} surveys. However, both of these analyses considered planet hosting probabilities at fixed planet mass instead of fixed mass ratio. Since lower mass ratio planets are more common \citep{suzuki2016exoplanet}, for mass ratios $\geq 10^{-4}$, the demonstration that higher mass hosts are more likely for a fixed mass ratio is stronger than the same claim at fixed planet mass. If the planet hosting probability was independent of host mass at fixed mass ratio, the hosting probability would be higher for higher mass hosts at fixed planet mass, since this implies a larger mass ratio, $q$, for the lower mass hosts (as long as $\geq 10^{-4}$).

\section{Conclusion}\label{sec:conclusion}
We have presented the discovery of MOA-2020-BLG-135Lb, a new Neptune-class planet uncovered by the light-curve analysis for the microlensing event MOA-2020-BLG-135. By applying the \citet{bennett2010efficient} process, our analysis has revealed a planet-host mass ratio of $q = 1.52_{-0.31}^{+0.39} \times 10^{-4}$, and separation $s\approx1$. 

To estimate the lens system properties for MOA-2020-BLG-135, we have conducted a Bayesian analysis using the \citet{bennett2014moa} Galactic model. When considering that all stars have equal probability of hosting a planet, using a mass function prior uniform in $M$, we have found a planet mass of $m_\mathrm{planet} = 11.3_{-6.9}^{+19.2} \;M_\oplus$, a host-star with mass of $M_\mathrm{host} = 0.23_{-0.14}^{+0.39} \;M_\odot$ and magnitude $K_\mathrm{L} = 22.2_{-2.3}^{+1.9}$, located at a distance $D_\mathrm{L} = 7.9_{-1.0}^{+1.0} \;\mathrm{kpc}$. Under the assumption that the hosting-probability scales in proportion to the stellar mass, we have estimated $m_\mathrm{planet} = 25_{-15}^{+22} \;M_\oplus$, $M_\mathrm{host} = 0.53_{-0.32}^{+0.42} \;M_\odot$, $K_\mathrm{L} = 20.4_{-2.3}^{+2.0}$, and $D_\mathrm{L} = 8.3_{-1.0}^{+0.9} \;\mathrm{kpc}$. 

The previous results from RV and direct imaging \citep{johnson2007new,johnson2010giant,nielsen2019gemini} considered planet hosting probabilities for fixed planet mass, whereas the MOA-II exoplanet microlens statistical analysis considers mass ratio. MOA-2020-BLG-135Lb is an important detection for completeness of the extended MOA-II exoplanet microlens statistical sample. Additionally, high-angular resolution follow-up observations for this event are certainly recommended in the future for restricting the mass values of the host star and its planet.

\section*{Acknowledgments}
We thank the government of New Zealand for their strict strategy against COVID-19, which gave MOA minimal down time during the pandemic. The observatory was kept open for most of the observing season, allowing us to observe this planetary event in 2020. We must also thank the KMTNet Collaboration for the valuable paper discussion, especially Jennifer Yee, Cheongho Han, and Andrew Gould. We would also like to thank Keming Zhang for his helpful conversation about his two papers.

Work by S.I.S., D.P.B., A.B., A.V., and G.R. was supported by NASA under award number 80GSFC21M0002. The MOA project is supported by JSPS KAK-ENHI Grant Number JSPS24253004, JSPS26247023, JSPS23340064, JSPS15H00781, JP16H06287,17H02871 and 19KK0082. Work by C.R. was supported by the Alexander von Humboldt Foundation, and was partly carried out within the framework of the ANR project COLD-WORLDS supported by the French \textit{Agence Nationale de la Recherche} with the reference ANR-18-CE31-0002. S.M. acknowledges support from JSPS KAKENHI Grant Number 21J11296. 
This research uses data obtained through the Telescope Access Program (TAP), which has been funded by the TAP member institutes. W.Zang, S.M. and W.Zhu acknowledge support by the National Science Foundation of China (Grant No. 12133005). W.Zhu acknowledge the science research grants from the China Manned Space Project with No.\ CMS-CSST-2021-A11. This work is partly based on observations obtained with MegaPrime/MegaCam, a joint project of CFHT and CEA/DAPNIA, at the Canada-France-Hawaii Telescope (CFHT) which is operated by the National Research Council (NRC) of Canada, the Institut National des Sciences de l'Univers of the French Centre National de la Recherche Scientifique (CNRS), and the University of Hawaii. This research is supported by Tsinghua University Initiative Scientific Research Program (Program ID 2019Z07L02017).

\software{Matplotlib \citep{matplotlib}, MOAna \citep{moana}, NumPy \citep{Oliphant2015}, SciPy \citep{scipy}.}

\bibliography{bibliography.bib}
\bibliographystyle{aasjournal}
\clearpage

\end{document}